\documentclass{llncs}

\newcommand{\when}{:\!\!-\;}
\newcommand{\comment}[1]{}
\newcommand{\condtext}[2]{#1}
\newcommand{\pdset}[1]{{\cal #1}}
\newcommand{\vct}[1]{\mbox{$\overline{#1}$}}
\newcommand{\staticcall}[1]{{#1}}
\newcommand{\staticarg}[1]{{#1}}
\newcommand{\dyncall}[1]{\underline{#1}}
\newcommand{\dynarg}[1]{\underline{#1}}

     {\begin{list}{$\bullet$}{
     \setlength{\itemsep}{0 pt}
     \setlength{\parsep}{0 pt}
     \setlength{\topsep}{0 pt} }}
     {\end{list}}
\newenvironment{pitemize}
     {\begin{list}{}{
     \setlength{\itemsep}{0 pt}
     \setlength{\parsep}{0 pt}
     \setlength{\topsep}{2 pt} }}
     {\end{list}}
\newenvironment{SProg}{\begin{small}\begin{tt}\begin{tabular}[c]{l}}{%
                       \end{tabular}\end{tt}\end{small}}

\begin{document}

\title{A Polyvariant Binding-Time Analysis for \\
           Off-line Partial Deduction}
\titlerunning{BTA for Off-line Partial Deduction}
\author{Maurice Bruynooghe \and Michael Leuschel
                           \and Konstantinos Sagonas}
\authorrunning{Maurice Bruynooghe et al.}
\tocauthor{Maurice Bruynooghe,  Michael Leuschel, Konstantinos Sagonas
  (Katholieke Universiteit Leuven)}
\institute{Katholieke Universiteit Leuven, Department of Computer Science\\
        Celestijnenlaan 200A, B-3001 Heverlee, Belgium\\
\email{ $\{$maurice,michael,kostis$\}$@cs.kuleuven.ac.be}}


\maketitle
\pagestyle{plain}


\condtext{\vspace*{-2pt}}{}
\begin{abstract}
  We study the notion of binding-time analysis for logic programs.
  We formalise the unfolding aspect of an on-line partial deduction
  system as a Prolog program.  Using abstract interpretation, we
  collect information about the run-time behaviour of the program.
  We use this information to make the control decisions about the
  unfolding at analysis time and to turn the on-line system into an
  off-line system.  We report on some initial experiments.
\end{abstract}


\section{Introduction} \label{sec:intro}
Partial evaluation and partial deduction are well-known techniques for
specialising respectively functional and logic programs.  While both
depart from the same basic concept, there is quite a divergence
between their application and overall approach.  In functional
programming, the most widespread approach is to use \emph{off-line
  specialisers}.  These are typically very simple and fast
specialisers which take (almost) no control decisions concerning the
degree of specialisation.  In this context, the specialisation is
performed as follows: First, a \emph{binding-time analysis} (BTA) is
performed on the program which annotates all its statements as either
``reducible'' or ``non-reducible''.
The annotated program is then
passed to the off-line specialiser, which executes the statements
marked reducible and produces residual code for the statements marked
non-reducible.  In logic programming, the \emph{on-line} approach is
almost the only one used.  All work is done by a complex on-line
specialiser which monitors the whole specialisation process and
decides on the degree of specialisation while specialising the
program.  A few researchers have explored off-line specialisation, but
lacking an appropriate notion of BTA, they worked with hand-annotated
programs, something which is far from being practical.  Until now, it
was unclear how to perform BTA for logic programs.

The current paper remedies this situation.  It develops a BTA for
logic programs, not by translating the corresponding notions from
functional programming to logic programming, but by departing from
first principles.  Given a logic program to be specialised, we develop
a logic program which performs its on-line specialisation.  The
behaviour of this program is analysed and the results are used to take
all decisions w.r.t.\ the degree of specialisation off-line.  This
turns the on-line specialiser into an off-line specialiser.  A
prototype has been built and the quality and speed of the off-line
specialisation has been evaluated.

\condtext{}{Section \ref{sec:back} recalls some background material and
the relevant
related work.  Section~\ref{sec:BTA} describes the on-line specialiser
and develops the BTA which allows to turn the on-line specialiser into
an off-line one.  Section~\ref{sec:impl} reports on the implementation
of the current prototype while Section~\ref{sec:exp+bench} evaluates
the resulting off-line specialiser.  Finally, Section~\ref{sec:conclu}
summarises the achievements and discusses future work.}

\section{Background} \label{sec:back}

\subsection{Partial Deduction}
In contrast to ordinary (full) evaluation, a {\em partial evaluator\/}
receives a program~$P$ along with only {\em part\/} of its input,
called the {\em static input}.  The remaining part of the input,
called the {\em dynamic input}, will only be known at some later
point in time.  Given the static input $S$, the partial evaluator then
produces a {\em specialised\/} version $P_S$ of $P$ which, when given
the dynamic input $D$, produces the same output as the original
program $P$.  The goal is to exploit the static input in order to
derive a more efficient program.

In the context of logic programming, full input to a program $P$
consists of a goal $G$ and evaluation corresponds to constructing a
complete SLDNF-tree for $P\cup\{G\}$.  The static input is given in
the form of a {\em partially instantiated\/} goal $G'$ (and the
specialised program should be correct \condtext{}{and more efficient}
for all \condtext{}{goals which are} instances of $G'$).

A technique which produces specialised programs is known under the
name of
\condtext{
\emph{partial deduction}~\cite{Komorowski@POPL-82}.
}{
\emph{partial deduction}.  Partial deduction originates from
\cite{Komorowski-thesis, Komorowski@POPL-82} (introductions can be
found in~\cite{Komorowski@META-92,Gallagher@PEPM-93,Leuschel-thesis}).
}
Its general idea is to construct a finite set of atoms $\pdset{A}$
and a finite set of finite, but possibly {\em incomplete\/} SLDNF-trees
(one for every%
\footnote{Formally, an SLDNF-tree is obtained from an atom or goal by
  what is called an {\em unfolding rule}.} atom in $\pdset{A}$) which
``cover'' the possibly infinite SLDNF-tree for $P\cup\{G'\}$.  The
derivation steps in these SLDNF-trees correspond to the computation
steps which have been performed beforehand 
\condtext{}{by the {\em  partial deducer\/}} 
and the specialised program is then extracted
from these trees by constructing one specialised clause per
non-failing branch.

In partial deduction one usually distinguishes two levels of control:
the {\em global control\/}, determining the set $\pdset{A}$, thus
deciding {\em which\/} atoms are to be partially deduced, and the {\em
  local control\/},
guiding construction of the finite 
SLDNF-trees for each individual atom in $\pdset{A}$ and thus
determining {\em what\/} the definitions for the partially deduced
atoms look like.

\subsection{Off-line vs.\ On-line Control}
The (global and local) control problems of partial evaluation and deduction
 in general have been tackled from
 two different angles: the so-called {\em on-line\/} versus
 {\em off-line\/} approaches.
The {\em on-line\/} approach performs
 all the control decisions
 {\em during\/} the actual specialisation phase.
The {\em off-line\/} approach on the other hand
 performs a (binding-time) analysis phase {\em prior\/} to
 the actual specialisation phase.
This analysis starts from a description of which parts of the inputs
 will be ``\emph{static}'' (i.e.\ sufficiently known)
 and provides \emph{binding-time annotations} which encode the control
 decisions to be made by the specialiser, so that the specialiser
 becomes much more simple and efficient.
 
Partial evaluation of functional programs
 \cite{ConselDanvy@POPL-93,PartEval-book} has mainly stressed off-line
 approaches, while supercompilation of functional
\condtext{
 \cite{Turchin@TOPLAS-86,SorensenGlueck@ILPS-95}
}{
 \cite{Turchin@TOPLAS-86,SorensenGlueck@ILPS-95,GlueckSorensen@PE-96}
}
 and partial deduction of logic programs
\condtext{
\cite{GallagherB@NGC-91,BruynoogheDeSchreyeMartens@NGC-92,Bol@JLP-93,%
Sahlin@NGC-93,MartensDeSchreye@JLP-96,LeuschelMartensDeSchreye@TOPLAS-98}%
}{
\cite{GallagherB@NGC-91,BruynoogheDeSchreyeMartens@NGC-92,Bol@JLP-93,%
Sahlin@NGC-93,Leuschel@LOPSTR-95,MartensDeSchreye@JLP-96,%
MartensGallagher@ICLP-95,LeuschelMartensDeSchreye@TOPLAS-98}
}
 have concentrated on on-line control.

On-line methods, usually obtain better specialisation, because no
control decisions have to be taken beforehand, i.e.\ at a point where
the full specialisation information is not yet available.  The main
reasons for using the off-line approach are to make specialisation
itself more efficient and, due to a simpler specialiser algorithm,
 enable effective self-application (specialisation of the specialiser)
 \cite{JonesSestoftSondergaard@LSC-89}.
\condtext{}{
 It also allows one to more easily develop hand-written compiler
 generators (see e.g.\ \cite{Romanenko:88,handcogen@PLILP-94,%
GlueckJoergensen:PLILP95} for functional programming and
 \cite{JorgensenLeuschel@PE-96,Leuschel-thesis} for logic programming),
 a technique which reaps (most) benefits of
 self-application without having to devise a self-applicable
 specialiser.}

Few authors discuss off-line specialisation in the context of
 logic programming
\condtext{
\cite{MogensenBondorf:LOPSTR92,JorgensenLeuschel@PE-96},
}{
\cite{MogensenBondorf:LOPSTR92,JorgensenLeuschel@PE-96,Leuschel-thesis},
}
 mainly because so far
 no automated binding-time analysers have been developed.
This paper aims to remedy this problem.

\section{Towards BTA for partial deduction} \label{sec:BTA}

\subsection{An on-line specialiser} \label{sec:on-line}
The basic idea of BTA in functional programming is to model the flow
of static input: the arguments of a function call flow to the function
body, the result of a function flows back to the call expression.  The
expressions are annotated {\em reducible\/} when enough of their parameters
are static, i.e.\ will be known at specialisation time, to allow the
(partial) computation of the expression. Modelling the dataflow gives
a system of inequalities over variables in a domain $\{static,dynamic\}$
whose least solution yields the best annotation.

This approach does not immediately translate to logic programs.
Problems are that the dataflow in unification is bidirectional and
that the degree of instantiation of a variable can change over its
lifetime (see also~\cite{JorgensenLeuschel@PE-96}).

We follow a different approach and reconstruct binding-time analysis
from first principles.  We start with a Prolog program which performs
the unfolding decisions of an on-line specialiser.  However, whereas
real on-line specialisers base their unfolding decisions on the
history of the specialisation phase, ours bases its decisions solely
on the actual arguments of the call (which can be more easily
approximated off-line).  This is in agreement with the off-line
specialisers for functional languages which base their decision to
evaluate or residualise an expression on the availability of the
parameters of that expression.  The next step will be to analyse the
behaviour of this program (the binding-time analysis) and to use the
results to make the unfolding decisions at compile time.

First we develop the on-line specialiser. 
Assuming that for each predicate $p/m$ a
\emph{test} predicate $\mathit{unfold}\_p/m$ exists which decides whether
to unfold a call or not, we obtain an on-line specialiser by replacing
each call $p(\vct{t})$ by
\[
    ( \mathit{unfold}\_p(\vct{t}) \rightarrow p(\vct{t}) ; 
    \mathit{memoise\_p}(\vct{t}) )
\]

A call to $\mathit{memoise\_p}(\vct{t})$ informs the specialiser that
the call $p(\vct{t})$ has to be residualised. The specialiser has to
check whether (a generalisation of) $p(\vct{t})$ has already been
specialised ---if not it has to initiate the specialisation of (a
generalisation of) $p(\vct{t})$--- and has to perform appropriate
renaming of predicates to ensure that residual code calls the proper
specialised version of the predicate it calls.
 
\begin{example}[Funny append]\label{ex:funnyapp}{\rm
 Consider the following on-line specialiser for a variant,
  $\mathit{funnyapp/3}$ of the $\mathit{append/3}$ predicate in
  which the first two arguments of the recursive call have been swapped:
  \begin{center}
  \begin{small}
  \fbox{\mbox{
  \begin{SProg}
    funnyapp([],X,X).                                           \\
    funnyapp([X|U],V,[X|W]) :-                                  \\
    \ \ \ \ ( unfold\_funnyapp(V,U,W) -> funnyapp(V,U,W)        \\
            \hspace{14em}         ;  memoise\_funnyapp(V,U,W) ). \\
    unfold\_funnyapp(X,Y,Z) :- ground(X). \\
  \end{SProg}
  }}
  \end{small}
  \end{center}
Specialising this program for a query {\tt funnyapp([a,b],L,R)}
results in the specialised clause (the residual call is renamed as
{\tt funnyapp\_1}) 
  \begin{pitemize}{\small
  \item {\tt funnyapp([a,b],L,[a|R1]) :- funnyapp\_1(L,[b],R1).}
  }\end{pitemize}
Specialising the {\tt funnyapp} program for the residual call {\tt
  funnyapp(L,[b],R1)} gives (after renaming) the clauses
  \begin{pitemize}{\small
  \item {\tt funnyapp\_1([],[b],[b]).}
  \item {\tt funnyapp\_1([X|U],[b],[X,b|R]) :- funnyapp\_2(U,[],R).}
  }\end{pitemize}
Once more specialising, now for the residual call {\tt
  funnyapp(U,[],R)}, gives 
  \begin{pitemize}{\small
  \item {\tt funnyapp\_2([],[],[]).}
  \item {\tt funnyapp\_2([X|U],[],[X|U]).} }\end{pitemize} 
This completes the specialisation.  Note that the sequence of residual
calls is terminating in this example.  In general, infinite sequences
are possible. They can be avoided by generalising some arguments of
the residual calls before specialising. 
} \end{example}

In the above example, instead of using {\tt ground(X)} as condition of
unfolding, one could also use the test {\tt nilterminated(X)}.%
This would allow to obtain the same level of specialisation for a query
{\tt funnyappend([X,Y],L,R)}.  This test is another example of a so
called \emph{rigid} or \emph{downward closed} property: if it holds
for a certain term, it holds also for all its instances.  Such
properties are well suited for analysis by means of abstract
interpretation.

\subsection{From on-line to off-line} \label{sec:analysis}

Turning the on-line specialiser into an off-line one requires to
determine the $\mathit{unfold}\_p/n$ predicates during a
preceding analysis and to decide on whether to replace the
$(\mathit{unfold}\_p(\vct{t}) \rightarrow p(\vct{t}); memoise(p(\vct{t})))$
construct either by $p(\vct{t})$ or by $memoise(p(\vct{t}))$.
The decision has to be based on a safe estimate of the calls
$\mathit{unfold}\_p(\vct{t})$
which will occur during the specialisation.
Computing such safe approximations is exactly the purpose of
\emph{abstract interpretation}~\cite{Cousot@POPL-77}.  

\begin{center}
\begin{small}
\fbox{\mbox{
\begin{SProg}
  :- $\{${\em grnd(L1)}$\}$ fap1(L1,L2,R) $\{${\em grnd(L1)}$\}$.  \\
  fap1([],X,X).                                         \\
  fap1([X|U],V,[X|W]) :- $\{${\em grnd(X,U)}$\}$        \\
  \hspace{2em}( unf\_fap1(V,U,W) $\{${\em grnd(X,U,V)}$\}$ -> \\
  \hspace{5em} $\{${\em grnd(X,U,V)}$\}$
                        fap2(V,U,W) $\{${\em grnd(X,U,V,W)}$\}$\\
  \hspace{5em}; $\{${\em grnd(X,U)}$\}$
                        memo\_fap3(V,U,W) $\{${\em grnd(X,U)}$\}$\\
  \hspace{2em}) $\{${\em grnd(X,U)}$\}$.\\ 
  unf\_fap1(X,Y,Z) :- $\{${\em grnd(Y)}$\}$ ground(X)
                      $\{${\em grnd(X,Y)}$\}$. \\
  memo\_fap3(X,Y,Z).
\end{SProg}
}}
\end{small}
\end{center}

By adding to the code of Example~\ref{ex:funnyapp} the fact
{\small\tt memoise\_funnyapp(X,Y,Z).}, and an appropriate handling
of the Prolog built-in {\tt ground/1}, one can run a goal-dependent
\emph{polyvariant} groundness analysis (using e.g.\ PLAI coupled
with the set-sharing domain) for a query where the first argument
is ground and obtain the above annotated program.  The annotated
code for the version {\tt fap2} is omitted because it is irrelevant
for us.  Indeed, inspecting the annotations for {\tt unf\_fap1} we
see that the analysis cannot infer the groundness of its first
argument.  So we decide off-line not to unfold, we cancel the test
and the {\em then} branch and simplify the code into:

\begin{center}
\begin{small}
\fbox{\mbox{
\begin{SProg}
  :- $\{${\em grnd(L1)}$\}$ fap1(L1,L2,R) $\{${\em grnd(L1)}$\}$. \\
  fap1([],X,X).                                                   \\
  fap1([X|U],V,[X|W]) :- $\{${\em grnd(X,U)}$\}$ memo\_fap3(V,U,W)
                         $\{${\em grnd(X,U)}$\}$.
\end{SProg}
}}
\end{small}
\end{center}

The residual call to {\tt funnyappend} has a different call pattern
than the original call: its second argument is now ground.  Thus we
perform a second analysis and obtain (the annotated code for {\tt
fap4} is omitted):

\begin{center}
\begin{small}
\fbox{\mbox{
\begin{SProg}
  :- $\{${\em grnd(L2)}$\}$ fap3(L1,L2,R) $\{${\em grnd(L2)}$\}$.  \\
  fap3([],X,X).                                         \\
  fap3([X|U],V,[X|W]) :-  $\{${\em grnd(V)}$\}$         \\
  \hspace{2em}( unf\_fap2(V,U,W) $\{${\em grnd(V)}$\}$ ->   \\
  \hspace{5.5em} $\{${\em grnd(V)}$\}$ fap4(V,U,W) $ \{${\em grnd(V)}$\}$\\
  \hspace{5em}; $\{${\em grnd(V)}$\}$ memo\_fap5(V,U,W)
                $\{${\em grnd(V)}$\}$\\
  \hspace{2em}) $\{${\em grnd(V)}$\}$.\\
  unf\_fap2(X,Y,Z) :- $\{${\em grnd(X)}$\}$ ground(X) $\{${\em grnd(X)}$\}$. \\
  memo\_fap5(X,Y,Z).
\end{SProg}
}}
\end{small}
\end{center}

This time, the annotations for {\tt unf\_fap2} show that the
groundness test will definitely succeed.  So we decide off-line always
to unfold and only keep the \emph{then} branch.  Moreover, the 
{\tt fap4} call has the same call pattern as the original call to
{\tt funnyapp}, so we also rename it as {\tt fap1}.  This yields
the second code fragment:

\begin{center}
\begin{small}
\fbox{\mbox{
\begin{SProg}
  :- $\{${\em grnd(L2)}$\}$ fap3(L1,L2,R) $\{${\em grnd(L2)}$\}$. \\
  fap3([],X,X).                                                   \\
  fap3([X|U],V,[X|W]) :-
        $\{${\em grnd(V)}$\}$ fap1(V,U,W)  $\{${\em grnd(V)}$\}$
\end{SProg}
}}
\end{small}
\end{center}

\condtext{\noindent}{}%
Applying the specialiser on these two code fragments for a query
 {\small\tt fap1([a,b],L,R)} gives the same specialised code as in
 Example~\ref{ex:funnyapp}. However, this time, no calls to {\tt
 unfold\_funnyapp} have to be evaluated during specialisation.

\subsection{Automation} \label{sec:automation}

To weave the step by step analysis sketched above in a single
analysis, a special purpose tool has to be built. We implemented a
system based on the abstract domain POS, also called
PROP~\cite{MarriottSondergaard@LOPLAS-93}.  It describes the state of
the program variables by means of \emph{positive} boolean formulas,
i.e., formulas built from $\leftrightarrow, \wedge$ and~$\vee$.  Its
most popular use is for groundness analysis.  In that case, the
formula $X$ expresses that the program variable $X$ is (definitely)
bound to a ground term, $X \leftrightarrow Y$ expresses that $X$ is
bound to a ground term iff $Y$ is, so an eventual binding of $X$ to a
ground term will be propagated to~$Y$.  This domain is extended with
$\mathit{false}$ as bottom element and is ordered by boolean
implication.  Groundness analysis corresponds to checking the
rigidity\footnote{A term is rigid w.r.t.\ a norm if all its instances
  have the same size w.r.t.\ the norm.} of program variables w.r.t.\ the
\emph{termsize norm}%
\footnote{The termsize norm~\cite{DecorteDeSchreye@ICLP-97} includes
  all subterms in the {\em measure} of the term.}
and abstracts a unification such as $X = [Y|Z]$ by the
boolean formula $X \leftrightarrow Y \wedge Z$.  However POS can also
be used with other semi-linear norms~\cite{CodishDemoen@SAS-94}.  In
e.g.\ normalised programs, it only requires to redefine the abstraction
of the unifications.  For example, with the \emph{listlength norm}%
\footnote{The listlength norm~\cite{DecorteDeSchreye@ICLP-97} includes
  only the tail of the list in the measure of the list (and measures
  other terms as 0); nil-terminated lists are rigid under this norm.},
unification of $X = [Y|Z]$ is abstracted as $X \leftrightarrow Z$,
and a formula $X$ means that the program variable $X$ is bound to a
term with a bounded listlength, i.e.\ either the term is a
nil-terminated list, or has a main functor which is not a list
constructor.
\condtext{}{
 Observe that this approach, which we will employ
 in Section~\ref{sec:exp+bench}, allows for what is called
 \emph{partially static} arguments in partial evaluation of
 functional programs (cf.\cite[Sect.~18.3]{PartEval-book}).
}

The analyser has to decide the outcome of the $\mathit{unfold}\_p$
test and has to decide which branch to take for further analysis while
doing the analysis.  Also it has to launch the analysis of the
generalisations of the memoised calls.  The generalisation we employ is
to replace an argument which is not rigid under the norm used in the
analysis by the abstraction of a fresh variable.  These requirements
exclude the direct use of the abstract compilation technique in the way
advocated by e.g.~\cite{CodishDemoen@JLP-95}.
One problem of the scheme of~\cite{CodishDemoen@JLP-95} is that it
handles constructs
        $(ground(X) \rightarrow p(\vct{t}); \mathit{memoise}(p(\vct{t})))$
too inaccurately.
The boolean formula is represented as a truth table, i.e.\ a set of
tuples, and the analyser processes the truth table a tuple at a time.
Therefore it cannot infer in a program point that $X$ is true, i.e.\ 
that $X$ is definitely ground, so it can never conclude that the {\em
else} branch cannot be taken.
 The other problem is that the analyses
 launched for the \emph{memoised} calls should not interfere 
  (i.e.\ output should not flow back) with the analysis
  of the clauses containing the memoised calls.
\condtext{Note}{
 To see why this is not easily done using the
 scheme of~\cite{CodishDemoen@JLP-95}, note
}
that defining $\mathit{memoise}$ as 
 $\mathit{memoise\_p}(X_1,\ldots,X_n) \when \mathit{copy}(X_1,Y_1), \ldots,
                                            \mathit{copy}(X_n,Y_n)$, 
 and abstracting $\mathit{copy}(X,Y)$ as $X \leftrightarrow Y$
 does not work:  The abstract success state of executing
 $\mathit{p}(Y_1,\ldots,Y_n)$ will update the abstractions
 of $X_1,\ldots,X_n$.

Our prototype binding-time analyser currently consists of $\approx 800$
lines of Prolog code and uses XSB~\cite{XSB@SIGMOD-94} as a generic
tool for semantic-based program analysis~\cite{CodishDemoenSagonas}.
The boolean formulas from the POS domain are represented by their
truth tables.  This representation enables abstract operations to have
straightforward implementations based on the projection and equi-join
operations of the relational algebra.
The disadvantage is that the size of truth tables quickly increases
with the number of variables in a clause.  The use of dedicated data
structures like BDDs\condtext{}{~\cite{BDDs@CS-92}} to represent the
boolean formulas as in~\cite{PropGAIA@JLP-95} often results in better
performance but at the expense of substantial programming efforts.

The main part of the analyser can be seen as a source-to-source
transformation (i.e.\ abstract compilation)
that given the program $P$ to be analysed, produces an
abstract program~$P^\alpha$ with suitable annotations. The abstract
program can be directly run under XSB (using tabling to ensure termination).
The execution leaves the results of the analysis in the XSB \emph{tables}.
\condtext{Each}{
Contrary to the scheme of~\cite{CodishDemoen@JLP-95}, that
manipulates boolean formulas of the POS domain a tuple at a time, each
}
predicate~$p/n$ of~$P$ is abstracted to a predicate~$p^\alpha/2$
whose arguments carry input and output \emph{sets} of tuples.
The core part of setting up the analysis is then to define the code
for the abstract interpretation of each call
(at a program point {\tt PP$_{\#}$} of interest):
\begin{equation}
   ( \mathit{unfold}\_p(\vct{X}) \rightarrow p(\vct{X})
                                        ; \mathit{memoise}\_p(\vct{X}) ).
\label{call_abstraction}
\end{equation}
is abstracted by the following code fragment:
\begin{pitemize}
\item {\tt $project($Args,TPP$_{in}$,TC$),$} 
\item {\tt ( $\mathit{unfold}\_p($TC$)$ ->
\item \hspace*{3.5em} $\mathit{unfold}($TC,PP$_{\#})$, $p^\alpha($TC,TR$)$}
\item \hspace*{3em} {\tt ;   TR=TC, $generalise($TC,TCG$),$
                           $memo($TCG,PP$_{\#}),$ $p^\alpha($TCG,\_$)$ ),}
\item {\tt $equi\_join($Args,TPP$_{in}$,TR,TPP$_{out}),$}
\end{pitemize}
Predicates $unfold/2$ and $memo/2$ which abstract the behaviour of
each call in the form of~(\ref{call_abstraction}) above
are \emph{tabled} predicates which have no effect on the computation,
but only record information containing the results of the analysis.
Their arguments are the current abstraction and the current program
point.  This information is then dumped from the XSB tables and is
fed to the off-line specialiser.
The variable {\tt TPP$_{in}$} holds the truth table which represents
the abstraction of the program state in the point prior to the call.
The call to $project/3$ projects the truth table on the positions
{\tt Args} of the variables $\vct{X}$ participating in the call. 
The result is {\tt TC} (Tuples of the Call).
The predicate $unfold\_p/1$ (currently supplied by the user for
each predicate~$p/n$ to be analysed) inspects {\tt TC} to decide
whether there is sufficient information to unfold the call.
If it succeeds the \emph{then}
branch is taken which analyses the effects of unfolding $p/n$.  This
is done by executing $p^\alpha/2$ with {\tt TC} as abstraction of
the call state.  The analysis returns {\tt TR} as abstraction of
the program state reached after unfolding~$p/n$.  If the call to
$unfold\_p/1$ fails, the call is memoised, and the program state
remains unchanged, so {\tt TR = TC}.  The generalisation of
the memoised call also needs to be analysed; therefore the {\em else}
branch first generalises the current state {\tt TC} into {\tt TCG}
by erasing all dependencies for non-rigid arguments%
\footnote{A position is rigid if it has an ``s'' in each tuple e.g.\
  {\tt $generalise($[p(s,s,s), p(s,d,d)],TCG$)$} 
          yields {\tt TCG = [p(s,s,s), p(s,s,d), p(s,d,s), p(s,d,d)]}.}
and then calls $p^\alpha/2$ with {\tt TCG} as initial state, but
takes care not to use the returned abstract state as the bindings
resulting from specialising memoised calls do not flow back.  These
actions effectively realise the intended functionality of
$\mathit{memoise\_p}/1$. 
Finally, the new program state {\tt TR} over the variables $\vct{X}$
has to be propagated to the other program variables described by the
initial state {\tt TPP$_{in}$}.  This is achieved simply by taking
the equi-join over the {\tt Args} of {\tt TPP$_{in}$} and {\tt TR}.
The new program state is described by {\tt TPP$_{out}$}.

One of our examples (see Section~\ref{sec:liftsolve}) uses two
different norms in the $\mathit{unfold}$ tests: the term norm which
tests for groundness, and the listlength norm which tests for the
boundedness of lists (whether lists are nil-terminated).
This does not pose a problem for our framework, we simply use a truth
table which encodes two boolean formulas, one for the term norm and
one for the listlength norm.

\section{Some Experiments and Benchmarks} \label{sec:exp+bench}

We first discuss the {\tt parser} and {\tt liftsolve} examples
\condtext{
 from~\cite{JorgensenLeuschel@PE-96}.
}{
 from~\cite{JorgensenLeuschel@PE-96,Leuschel-thesis}.
}

\subsection{The {\tt parser} example}
A small generic parser for languages
 defined by grammars of the form $S ::= aS | X$
 ($X$ is a placeholder for a terminal symbol
  as well as the first argument to {\tt nont/3}; arguments
  2 and 3 represent the string to be parsed as a difference list):
\begin{pitemize}\begin{footnotesize}
\item {\tt nont(X,T,R) :- t(a,T,V),nont(X,V,R).}
\item {\tt nont(X,T,R) :- t(X,T,R).}
\item {\tt t(X,[X|Es],Es).}
\end{footnotesize}\end{pitemize}
A termination analysis can easily determine that
 calls to {\tt t/3} always terminate and that
 calls to {\tt nont/3} terminate if their second argument
 is ground.
One can therefore derive the following
 unfold predicates:
\begin{pitemize}\begin{footnotesize}
\item {\tt unfold\_t(X,S1,S2).}
\item {\tt unfold\_nont(X,T,R) :- ground(T).}
\end{footnotesize}\end{pitemize}
Performing our analysis for
 the entry point {\small {\tt :- } $\{grnd(X)\}$ {\tt nont(X,\_,\_)}}
 we obtain
 the following annotated program (dynamic arguments [i.e.\ non-ground ones]
 and non-reducible predicates [i.e.\ memoised ones]
 are \dyncall{underlined}):
\begin{pitemize}\begin{footnotesize}
\item {\tt nont(\staticarg{X},\dynarg{T},\dynarg{R}) :-
         \staticcall{t}(\staticarg{a},\dynarg{T},\dynarg{V}),
                \dyncall{nont}(\staticarg{X},\dynarg{V},\dynarg{R}).}
\item {\tt nont(\staticarg{X},\dynarg{T},\dynarg{R}) :- 
        \staticcall{t}(\staticarg{X},\dynarg{T},\dynarg{R}).}
\item {\tt t(\staticarg{X},\dynarg{[X|Es]},\dynarg{Es}).}
\end{footnotesize}\end{pitemize}
Feeding this information into the off-line system {\sc logen}
 \cite{JorgensenLeuschel@PE-96}
and specialising
  {\tt\small nont(\staticarg{c},\dynarg{T},\dynarg{R})}, we obtain:
\begin{pitemize}\begin{footnotesize}
\item     nont\_\_0([a$|$B],C) :- nont\_\_0(B,C).
\item     nont\_\_0([c$|$D],D).
\end{footnotesize}\end{pitemize}
Analysing the same specialiser for
 {\small {\tt :- } $\{grnd(T)\}$ {\tt nont(\_,\staticarg{T},\_)}} yields:
\begin{pitemize}\begin{footnotesize}
\item {\tt nont(\dynarg{X},\staticarg{T},\dynarg{R}) :-
  \staticcall{t}(\staticarg{a},\staticarg{T},\dynarg{V}),
                \staticcall{nont}(\dynarg{X},\staticarg{V},\dynarg{R}).}
\item {\tt nont(\dynarg{X},\staticarg{T},\dynarg{R}) :-
  \staticcall{t}(\dynarg{X},\staticarg{T},\dynarg{R}).}
\item {\tt t(\dynarg{X},\staticarg{[X|Es]},\dynarg{Es}).}
\end{footnotesize}\end{pitemize}
Feeding this information into {\sc logen}
 and specialising
 {\tt\small nont(\dynarg{X},\staticarg{[a,a,c]},\dynarg{R})}
 yields:
\begin{pitemize}\begin{footnotesize}
\item      nont\_\_0(c,[]).
\item      nont\_\_0(a,[c]).
\item      nont\_\_0(a,[a,c]).
\end{footnotesize}\end{pitemize}

\subsection{The {\tt liftsolve} example} \label{sec:liftsolve}

The following program is a meta-interpreter 
 for the ground representation, in which the goals are
 ``lifted'' to the non-ground representation for resolution.
To perform the lifting, an accumulating parameter is used to keep track
 of the variables that have already been encountered and generated.
The predicate {\tt\small mng} and {\tt\small l\_mng}
 transform (a list of) ground
  terms (the first argument) into (a list of) non-ground terms
 (the second argument; the third and fourth arguments represent the
 incoming and outgoing accumulator respectively).
The predicate {\tt\small solve} uses these predicates
 to ``lift'' clauses of a
 program in ground representation (its first argument) and then
 use them for resolution with a non-ground goal (its second argument)
 to be solved.\\
\begin{SProg}
 solve(GrP,[]). \\
solve(GrP,[NgH|NgT]) :-\\
        \hspace*{1cm} non\_ground\_member(term(clause,[NgH$|$NgBdy]),GrP),\\
        \hspace*{1cm} solve(GrP,NgBdy), solve(GrP,NgT). \\
non\_ground\_member(NgX,[GrH$|$\_GrT]) :-
        make\_non\_ground(GrH,NgX).     \\
non\_ground\_member(NgX,[\_GrH$|$GrT]) :-
        non\_ground\_member(NgX,GrT).   \\
make\_non\_ground(G,NG) :- mng(G,NG,[],\_Sub). \\
mng(var(N),X,[],[sub(N,X)]).   \\
 mng(var(N),X,[sub(N,X)$|$T],[sub(N,X)$|$T]).  \\
 mng(var(N),X,[sub(M,Y)$|$T],[sub(M,Y)$|$T1]) :-
        N \verb|\==| M, mng(var(N),X,T,T1).    \\
 mng(term(F,Args),term(F,IArgs),InS,OutS) :-
        lmng(Args,IArgs,InS,OutS).   \\
 lmng([],[],Sub,Sub).        \\
 lmng([H|T],[IH$|$IT],InS,OutS) :-   \\
        \hspace*{1cm} mng(H,IH,InS,InS1), lmng(T,IT,InS1,OutS).
\end{SProg}

\noindent
The following unfold predicates can be
 derived by a termination analysis:
\begin{pitemize}\begin{footnotesize}\begin{tt}
\item unfold\_lmng(Gs,NGs,InSub,OutSub) :- 
        ground(Gs), bounded\_list(InSub).
\item unfold\_mng(G,NG,InSub,OutSub) :-   
        ground(G), bounded\_list(InSub).
\item unfold\_make\_non\_ground(G,NG) :- ground(G).
\item unfold\_non\_ground\_member(NgX,L) :- ground(L).
\item unfold\_solve(GrP,Query) :- ground(GrP).
\end{tt}\end{footnotesize}\end{pitemize}

\noindent
Analysing the specialiser
 for the entry point {\tt\small solve(\staticarg{ground},\_)} we obtain:

\begin{pitemize}\begin{small}
\item solve(\staticarg{GrP},\dynarg{[]}).
\item solve(\staticarg{GrP},\dynarg{[NgH$|$NgT]}) :-\\
      \hspace*{1cm}
      \staticcall{non\_ground\_member}(\dynarg{term(clause,[NgH$|$NgBdy])},
      \staticarg{GrP}),\\
       \hspace*{1cm} \dyncall{solve}(\staticarg{GrP},\dynarg{NgBdy}),
        \dyncall{solve}(\staticarg{GrP},\dynarg{NgT}).
\item non\_ground\_member(\dynarg{NgX},\staticarg{[GrH$|$\_GrT]}) :-
        \staticcall{make\_non\_ground}(\staticarg{GrH},\dynarg{NgX}).
\item non\_ground\_member(\dynarg{NgX},\staticarg{[\_GrH$|$GrT]}) :-
        \staticcall{non\_ground\_member}(\dynarg{NgX},\staticarg{GrT}).
\item make\_non\_ground(\staticarg{G},\dynarg{NG}) :-
        \staticcall{mng}(\staticarg{G},\dynarg{NG},
                         \staticarg{[]},\dynarg{\_Sub}).
\item mng(\staticarg{var(N)},\dynarg{X},\staticarg{[]},\dynarg{[sub(N,X)]}).
\item mng(\staticarg{var(N)},\dynarg{X},\staticarg{[sub(N,X)$|$T]},
                        \dynarg{[sub(N,X)$|$T]}).
\item mng(\staticarg{var(N)},\dynarg{X},\staticarg{[sub(M,Y)$|$T]},
                        \dynarg{[sub(M,Y)$|$T1]}) :-
    \staticarg{N} \verb|\==| \staticarg{M},
    \staticcall{mng}(\staticarg{var(N)},\dynarg{X},\staticarg{T},\dynarg{T1}).
\item mng(\staticarg{term(F,Args)},\dynarg{term(F,IArgs)},
                \staticarg{InS},\dynarg{OutS}) :-
        \staticcall{lmng}(\staticarg{Args},\dynarg{IArgs},
                             \staticarg{InS},\dynarg{OutS}).
\item lmng(\staticarg{[]},\dynarg{[]},\staticarg{Sub},\dynarg{Sub}).
\item lmng(\staticarg{[H$|$T]},\dynarg{[IH$|$IT]},
              \staticarg{InS},\dynarg{OutS}) :-
        \staticcall{mng}(\staticarg{H},\dynarg{IH},\staticarg{InS},\dynarg{InS1}),
        \fbox{\dyncall{lmng1}(\staticarg{T},\dynarg{IT},
                         {\bf \dynarg{InS1}},\dynarg{OutS})}.  
\item lmng1(\staticarg{[]},\dynarg{[]},\dynarg{Sub},\dynarg{Sub)}.
\item lmng1(\staticarg{[H$|$T]},\dynarg{[IH$|$IT]},
                 \dynarg{InS},\dynarg{OutS}) :-
        \dyncall{mng1}(\staticarg{H},\dynarg{IH},\dynarg{InS},\dynarg{InS1}),
        \dyncall{lmng1}(\staticarg{T},\dynarg{IT},
                             \dynarg{InS1},\dynarg{OutS}).
\item mng1(\staticarg{var(N)},\dynarg{X},\dynarg{[]},\dynarg{[sub(N,X)]}).
\item mng1(\staticarg{var(N)},\dynarg{X},
                \dynarg{[sub(N,X)$|$T]},\dynarg{[sub(N,X)$|$T]}).
\item mng1(\staticarg{var(N)},\dynarg{X},
                \dynarg{[sub(M,Y)$|$T]},\dynarg{[sub(M,Y)$|$T1]}) :-
        \staticarg{N} \dyncall{$\backslash$==} \dynarg{M},
        \dyncall{mng1}(\staticarg{var(N)},\dynarg{X},\dynarg{T},\dynarg{T1}).
\item mng1(\staticarg{term(F,Args)},\dynarg{term(F,IArgs)},
           \dynarg{InS},\dynarg{OutS}) :-
        \dyncall{lmng1}(\staticarg{Args},\dynarg{IArgs},
                             \dynarg{InS},\dynarg{OutS}).
\end{small}\end{pitemize}

One can observe that the call
 {\tt\small \dyncall{lmng1}(\staticarg{T},\dynarg{IT},
                         \dynarg{InS1},\dynarg{OutS})}
 has not been unfolded.
Indeed, the third argument {\small\tt InS1} is considered to be
 dynamic (non-ground)
 and the call to {\tt\small unfold\_lmng} will thus not always
 succeed.
However, based on the termination analysis, it is actually sufficient
 for termination if the third arguments to {\tt\small mng} and
 {\tt\small lmng} are bounded lists (as the listlength norm
 can be used in the termination proof).
If we use our prototype to also keep track of bounded lists
 we obtain the desired result:
 the call
 {\tt\small lmng1(T,IT,InS1,OutS)} can be unfolded
 as the first argument is ground and
 third argument can be inferred to be a bounded list.
By feeding the so obtained annotations into {\sc logen}
 \cite{JorgensenLeuschel@PE-96}
 we obtain a specialiser which removes (most of) the
 meta-interpretation overhead.
E.g.\ specialising
\begin{pitemize}\begin{tt}\begin{footnotesize}
\item  solve([term(clause,[term(q,[var(1)]), term(p,[var(1)])]),
\item  \hspace{2em}   term(clause,[term(p,[term(a,[])])])],G)
\end{footnotesize}\end{tt}\end{pitemize}
yields the following residual program:
\begin{pitemize}\begin{tt}\begin{footnotesize}
\item      solve\_\_0([]).
\item      solve\_\_0([term(q,[B])|C]) :- solve\_\_0([term(p,[B])]),solve\_\_0(C).
\item      solve\_\_0([term(p,[term(a,[])])|D]) :- solve\_\_0([]),solve\_\_0(D).
\end{footnotesize}\end{tt}\end{pitemize}

\subsection{Some Benchmarks}

We now study the efficiency and quality of our approach on a
 set of benchmarks.
Except for the {\tt parser} benchmark 
 all benchmarks come from the {\sc dppd} benchmark
 library \cite{Leuschel96:ecce-dppd}.
We ran our prototype analyser, {\sc bta}, that performs binding-time
 analysis and fed the result into the off-line compiler generator
 {\sc logen} \cite{JorgensenLeuschel@PE-96}
 in order to derive a specialiser for the task at hand.
The {\sc ecce} on-line partial deduction
 system \cite{Leuschel96:ecce-dppd}
  has been used for comparison
  (settings are the same as for {\sc ecce-x} in%
\condtext{
   \cite{LeuschelMartensDeSchreye@TOPLAS-98},
}{
   \cite{Leuschel-thesis,LeuschelMartensDeSchreye@TOPLAS-98},
}
   i.e.\ a mixtus like unfolding, a global control based upon 
   characteristic trees but no use of conjunctive partial deduction).
The interested reader can consult
\condtext{
 \cite{LeuschelMartensDeSchreye@TOPLAS-98}
}{
 \cite{Leuschel-thesis,LeuschelMartensDeSchreye@TOPLAS-98}
}
 to see how {\sc ecce} compares with other systems.
  
All experiments were conducted on a Sun Ultra-1 running SunOS~5.5.1.
{\sc ecce} and {\sc logen}
\condtext{were run}{and the original and specialised programs were all run}
using Prolog by BIM~4.1.0.
{\sc bta} was run on XSB~1.7.2.

\begin{table}[ht]
\condtext{\vspace*{-10pt}}{}
\begin{center}\begin{small}
\begin{tabular}{|l||c|ccc|c|} \hline
Benchmark &    {\sc ecce} - PD       & {\sc bta} & {\sc logen} & PD  & Ratio\\
\hline\hline
{\tt depth.lam     } &  0.34 s & 0.05 + 0.579 s  & 0.05 s& 0.003 s& 113\\
{\tt liftsolve.app } &  1.00 s & 0.079* + 1.841 s& 0.05 s& 0.006 s& 167\\
{\tt liftsolve.app4} & 12.32 s & $''$            & $''$  & 0.014 s& 880\\
{\tt match.kmp     } &  0.18 s & 0.06 + 0.031 s  & 0.01 s& 0.006 s& 30\\
{\tt parser        } &  0.06 s & 0.03 + 0.01 s   & 0.02 s& 0.001 s& 60\\
{\tt regexp.r1     } &  0.17 s & 0.039 + 0.031 s & 0.06 s& 0.006 s& 28\\
\hline
\end{tabular}
\end{small}
\caption{Analysis and Specialisation Times\label{table:benchmarks-tt}}
\end{center}
\condtext{\vspace*{-20pt}}{}
\end{table}

In Table~\ref{table:benchmarks-tt} one can see a summary of the
 transformation times.
The columns under {\sc bta} 
 contain:  the time to abstract and compile the program + the time for
 execution of the abstracted program (both under XSB). The column under
 {\sc logen} contains the time to generate the specialiser with
 {\sc logen} using the so obtained annotations. 
Observe, that for any given initial annotation, this has only
 to be performed {\em once}: the so obtained specialiser can
  then be used over and over again for different specialisation
  tasks.
E.g.\ the same specialiser was used for the
 {\tt liftsolve.app} and {\tt liftsolve.app4} benchmark.
The `*' for {\tt liftsolve.app} indicates the time for
 the abstract compilation only producing code for
 the groundness analysis.
The extra arguments and instructions for the bounded list analysis
 were added by hand (but will be generated automatically in the
 next version of the prototype).
 The column under {\sc
   PD} gives the time for the off-line specialisation.
 The last column of the table contains the ratio of running {\sc ecce}
 over running the specialisers generated by {\sc bta + logen}.
As can be seen, the specialisers produced by
 {\sc bta + logen} run 28 -- 880 times faster than {\sc ecce}.
We conjecture that for larger programs (e.g\ liftsolve
 with a very big object program) this difference can get even bigger.
Also, for 3 benchmarks the combined time of running
 {\sc bta + logen} and then the so obtained specialiser
 was less than running {\sc ecce},
 i.e.\ our off-line approach fares well even in
 ``one-shot'' situations.
Of course, to arrive at a fully automatic (terminating) system
 one will still have to add the time for the termination analysis,
 needed to derive the ``unfold'' predicates.

\begin{table}[ht]
\condtext{\vspace*{-15pt}}{}
\begin{center}\begin{small}
\begin{tabular}{|l||c|c|c|} \hline
Benchmark            & Original &{\sc ecce} & {\sc bta + logen} \\
\hline\hline
{\tt depth.lam     } &  0.08 s  &   0.00 s  &   0.06 s    \\
     ~               &    1     &$\approx$ 32 & 1.33      \\
{\tt liftsolve.app } &  0.13 s  &   0.01 s  &   0.01 s    \\
     ~               &    1     &    13     &    13       \\
{\tt liftsolve.app4} &  0.17 s  &   0.00 s  &   0.02 s    \\
     ~               &    1     &  $>$ 34   &    8.5      \\
{\tt match.kmp     } &  0.58 s  &   0.34 s  &   0.51 s    \\
     ~               &    1     &    1.71   &   1.14      \\
{\tt parser        } &  0.20 s  &   0.12 s  &   0.12 s    \\
     ~               &    1     &    1.74   &   1.74      \\
{\tt regexp.r1     } &  0.29 s  &   0.10 s  &   0.20 s    \\
     ~               &    1     &    2.9    &   1.5       \\
\hline
\end{tabular}
\end{small}
\caption{Absolute Runtimes and Speedups\label{table:benchmarks-su}}
\end{center}
\condtext{\vspace*{-25pt}}{}
\end{table}

Table~\ref{table:benchmarks-su} compares the efficiency of
 the specialised programs
 (for the run time queries see \cite{Leuschel96:ecce-dppd};
  for the {\tt parser} example we ran
\condtext{
{\small $\mathit{nont}(c,[a^{17},c,b],[b])$}
}{\\
{\small $\mathit{nont}(c,[a,a,a,a,a,a,a,a,a,a,a,a,a,a,a,a,a,a,c,b],[b])$}
}
   100 times).
As was to be expected,
 the programs generated by the on-line specialiser {\sc ecce}
 outperform those generated by our off-line system.
E.g.\ for the {\tt match.kmp} benchmark {\sc ecce} is able to derive
 a Knuth-Morris-Pratt style searcher, while off-line systems
 (so far) are unable to achieve such a feat.
However, one can see that the specialised programs generated
 by {\sc bta + logen} are still very satisfactory.
The most satisfactory application is
 {\tt liftsolve.app} (as well as {\tt liftsolve.app4}),
  where the specialiser generated
  by {\sc bta + logen} runs 167 (resp.\ 880) times faster than {\sc ecce}
  while producing residual code of equal (resp.\ almost equal) efficiency.
In fact, the specialiser
 compiled the append object program from the ground representation
 into the non-ground one in just 0.006 s
 (to be compared with e.g.\ the compilers generated by {\sc sage}
  \cite{Gurr:PHD} which run in the order of minutes).
Furthermore, the time to produce the residual program
 and then running it is less than the time needed to run the original
 program for the given set of runtime queries.
This nicely illustrates the potential of our approach for applications
 such as runtime code generation, where the specialisation time is
 (also) of prime importance.

\section{Discussion} \label{sec:discuss}
We have formulated a binding-time analysis for logic programs, and
have reported on a prototype implementation and on an evaluation of
its effectiveness.  To develop the binding-time analysis, we have
followed an original approach: Given a program $P$ to be analysed we
transform it into an on-line specialiser
 program $P'$, in which the unfolding decision are
explicitly coded as calls to predicates {\tt unfold\_p}. 
The on-line
specialiser is different from usual ones in the sense that it --- like
off-line specialisers --- uses the availability of arguments to decide
on the unfolding of calls.  Next, we apply abstract interpretation
---a binding-time analysis--- to gather information about the run-time
behaviour of $P'$. The information in the program points related to
{\tt unfold\_p} allows to decide whether the test will definitely
succeed ---in which case the unfolding branch is retained--- or will
possibly fail ---in which case the branch yielding residual code is
retained.  The resulting program 
 now behaves as an off-line specialiser as all unfolding
decisions have been taken at analysis time.

An issue to be discussed in more detail is the termination of the
specialisation.  First, a specialiser has a global control
component.  It must ensure that only a finite number of atoms are
specialised.  In our prototype, we generalise the residual calls
before generating a specialised version: arguments which are not
rigid%
\footnote{I.e., ``{\em static}'' from functional programming
 becomes ``{\em rigid} w.r.t.\ a given norm.''}
 w.r.t.\ the norm used in the unfolding condition are replaced
by fresh variables. This works well in practice but is not a
sufficient condition for termination.
In principle one could define the {\tt memoise\_p} predicates as:
\begin{pitemize}
\item {\tt memoise\_p(\vct{X}) :-
                copy\_term(\vct{X},\vct{Y}),
                generalise(\vct{Y},\vct{Z}), p(\vct{Z}).}
\end{pitemize}
and then generalise such that quasi-termination \cite{TabledUnfold@LOPSTR-97}
 of the program, where calls to {\tt p} are tabulated, can be
proven.  In practice,
the built-in {\tt copy\_term/2} and the built-ins needed to
implement {\tt generalise/2} will make this a non-trivial task.
Secondly, there is the local control component. It must ensure that
the unfolding of a particular atom terminates.  This is decided by
the code of the transformed program.  Defining the {\tt unfold\_p}
predicates by hand is error-prone and consequently not entirely
reliable.  In principle, one could replace the calls {\tt memoise\_p}
by {\tt true} and apply off-the-shelf tools for proving termination of
logic programs~\cite{LindenstraussSagiv@ICLP-97,CodishTaboch@ALP-97}.
Whether these will do well depends on how well they handle the
$\mathit{if-then-else}$ construct used in deciding on the unfolding
and the built-ins used in the rigidity test (e.g.\ the analysis has
to infer that {\tt X} is bounded and rigid w.r.t.\ the norm in the
program point following a test {\tt ground(X)}).  It is likely that
small extensions to these tools will suffice to apply them successfully
in proving termination of the unfolding\footnote{After a small extension
  by its author, the system of~\cite{LindenstraussSagiv@ICLP-97} could
  handle small examples.  However, so far we have not done exhaustive
  testing.},
at least when the unfolding conditions are based on rigidity tests
with respect to the norms used by those termination analysis tools. 

A more interesting approach for the local control problem is to
automatically generate unfolding conditions by program analysis.
Actually, one could apply a more general scheme for handling the
unfolding than the one used so far.  Having for each predicate
{\tt p/n} the original clauses with head {\tt p/n} and transformed
clauses with head {\tt pt/n}, the transformed clauses could be
derived from the original by replacing each call {\tt q/m} by:
\begin{pitemize}
\item {\tt ( terminates\_q(\vct{t}) -> q(\vct{t})\\
        \hspace*{2em}
      ; ( unfold\_q(\vct{t}) -> qt(\vct{t}) ; memoise\_q(\vct{t}) ) ) }
\end{pitemize}
In \cite{DecorteDeSchreye}, Decorte and De Schreye describe how the
constraint-based termination analysis of~\cite{DecorteDeSchreye@ICLP-97}
can be adapted to generate a finite set of ``most general'' termination
conditions (e.g.\ for {\tt append/3} they would generate rigidity
w.r.t.\ the listlength norm of the first argument and rigidity w.r.t.\
the listlength norm of the third argument as the two most general
termination conditions; for our {\tt funnyapp/3} they would generate
rigidity of the first and second argument w.r.t.\ the listlength norm as
the most general termination condition.). These conditions can be used
to define the {\tt terminates\_q} predicates. If they succeed, the
call {\tt q(\vct{t})} can be executed with the original code and is
guaranteed to terminate. Moreover, as they are based on rigidity, they
are very well suited to be approximated by our binding-time analysis.
Actually, in all our benchmarks programs, we were using termination
conditions for controlling the unfolding, so in fact we could have
further improved the speed of the specialiser by not checking the
condition on each iteration but using the above scheme.

Generating {\tt unfold\_q} definitions is a harder problem. It is
related to the generation of ``safe'' (i.e.\ termination ensuring)
delay declarations in languages such as MU-Prolog and G\"odel.
This is a subtle problem as discussed in
\cite{Naish92,MarchioriTeusink@ILPS-95}. For example, the condition
{\tt (nonvar(X); nonvar(Z))} is not safe for a call {\tt append(X,Y,Z)};
execution, and in our case unfolding, could go on infinitely for some
non-linear calls (e.g.\ {\tt append([a|L],Y,L)}).  Also the condition
{\tt nonvar/1} is not rigid.  (For funnyapp/3 we had rigid conditions,
however this is rather the exception than the rule.)  A safe unfolding
condition for {\tt append(X,Y,Z)} is
        {\tt linear(append(X,Y,Z)), (nonvar(X); nonvar(Z))}.
Linearity is well suited for analysis (e.g.~\cite{AbsUni:vol1000}),
but a test {\tt nonvar(X)} is not.  Moreover, unless {\tt X} is ground,
the test is typically not invariant over the different iterations of a
recursive predicate.  A solution could be to switch to a hybrid
specialiser: deciding the linearity test at analysis-time and the
simple {\tt nonvar} tests at run-time.  But as said above, perhaps
due to lack of a good application (for languages with delay, speed
is more important than safety), there seems to be no work on
generating such conditions.

Another hybrid approach is taken in a recent work independent of
ours~\cite{Martin}. This work also starts from the termination
condition. When it is violated, the size of the term w.r.t.\ the norm
used in the termination condition and the maximal reduction of the
size in a single iteration is used to compute the number of unfolding
steps. The program is transformed and calls to be unfolded are given
an extra argument initialised with the allowed number of unfolding
steps. An on-line test checks the value of the counter and the call is
residualised when the counter reaches zero.

\begin{small}
\subsection*{Acknowledgements}
M.\ Bruynooghe and M.\ Leuschel are supported by the Fund
 for Scientific Research - Flanders Belgium (FWO).
K.\ Sagonas is supported by the Research Council of the
 K.U.\ Leuven.
Some of the present ideas originated from discussions
 and joint work with Jesper J{\o}rgensen, and from the PhD.\ work of
 Dirk Dussart~\cite{Dussart-thesis}, to both of whom we are very
 grateful.
We thank Bart Demoen, Stefaan Decorte, Bern Martens, Danny De Schreye
 and Sandro Etalle
 for interesting discussions, ideas and comments.
\end{small}


\begin{small}

\vspace*{-10pt}
\begin{thebibliography}{10}
\vspace*{-5pt}

\bibitem{Bol@JLP-93}
R.~Bol.
\newblock {Loop Checking in Partial Deduction}.
\newblock {\em The Journal of Logic Programming}, 16(1\&2):25--46, May 1993.

\bibitem{AbsUni:vol1000}
M.~Bruynooghe, M.~Codish, and A.~Mulkers.
\newblock Abstracting unification: a key step in the design of logic program
  analyses.
\newblock In {\em Computer Science Today}, pages 406--442. Springer-Verlag,
  LNCS Vol. 1000, 1995.

\bibitem{BruynoogheDeSchreyeMartens@NGC-92}
M.~Bruynooghe, D.~De~Schreye, and B.~Martens.
\newblock {A General Criterion for Avoiding Infinite Unfolding During Partial
  Deduction}.
\newblock {\em New Generation Computing}, 11(1):47--79, 1992.

\bibitem{CodishDemoen@SAS-94}
M.~Codish and B.~Demoen.
\newblock {Deriving Polymorphic Type Dependencies for Logic Programs Using
  Multiple Incarnations of Prop}.
\newblock In B.~{Le Charlier}, editor, {\em Proceedings of the First
  International Symposium on Static Analysis}, number 864 in LNCS, pages
  281--297, Namur, Belgium, September 1994. Springer-Verlag.

\bibitem{CodishDemoen@JLP-95}
M.~Codish and B.~Demoen.
\newblock {Analysing Logic Programs using ``Prop''-ositional Logic Programs and
  a Magic Wand}.
\newblock {\em Journal of Logic Programming}, 25(3):249--274, December 1995.

\bibitem{CodishDemoenSagonas}
M.~Codish, B.~Demoen, and K.~Sagonas.
\newblock {Semantic-Based Program Analysis for Logic-Based Languages using
  XSB}.
\newblock K.U. Leuven TR CW 245. December 1996.

\bibitem{CodishTaboch@ALP-97}
M.~Codish and C.~Taboch.
\newblock {A Semantic Basis for Termination Analysis of Logic Programs and its
  Realization using Symbolic Norm Constraints}.
\newblock In {\em Proceedings of the Sixth International Conference on
  Algebraic and Logic Programming}, number 1298 in LNCS, pages 31--45.
  Springer-Verlag, September 1997.

\bibitem{ConselDanvy@POPL-93}
C.~Consel and O.~Danvy.
\newblock {Tutorial Notes on Partial Evaluation}.
\newblock In {\em Proceedings of the ACM Conference on Principles of
  Programming Languages}, pages 493--501, Charleston, South Carolina, January
  1993. ACM Press.

\bibitem{Cousot@POPL-77}
P.~Cousot and R.~Cousot.
\newblock {Abstract Interpretation: A Unified Lattice Model for Static Analysis
  of Programs by Construction or Approximation of Fixpoints}.
\newblock In {\em Conference Record of the Fourth ACM Symposium on Principles
  of Programming Languages}, pages 238--252, Los Angeles, California, January
  1977. ACM.

\bibitem{DecorteDeSchreye}
S.~Decorte and D.~{De Schreye}.
\newblock {Termination Analysis: Some Practical Properties of the Norm and
  Level Mapping Space}.
\newblock TR, Dept. Comp. Science, K.U. Leuven.

\bibitem{DecorteDeSchreye@ICLP-97}
S.~Decorte and D.~{De Schreye}.
\newblock {Demand-driven and Constraint-based Automatic Termination Analysis
  for Logic Programs}.
\newblock In L.~Naish, editor, {\em Proceedings of the Fourteenth International
  Conference on Logic Programming}, pages 78--92, Leuven, Belgium, July 1997.
  The MIT Press.

\bibitem{Dussart-thesis}
D.~Dussart.
\newblock {\em {Topics in Program Specialisation and Analysis for Statically
  Typed Functional Languages}}.
\newblock PhD thesis, Katholieke Universiteit Leuven, May 1997.

\bibitem{GallagherB@NGC-91}
J.~Gallagher and M.~Bruynooghe.
\newblock {The Derivation of an Algorithm for Program Specialisation}.
\newblock {\em New Generation Computing}, 9(3,4):305--333, 1991.

\bibitem{Gurr:PHD}
C.~A. Gurr.
\newblock {\em A Self-Applicable Partial Evaluator for the Logic Programming
  Language G\"odel}.
\newblock PhD thesis, Department of Computer Science, University of Bristol.

\bibitem{PartEval-book}
N.~D. Jones, C.~K. Gomard, and P.~Sestoft.
\newblock {\em {Partial Evaluation and Automatic Program Generation}}.
\newblock Prentice Hall International Series in Computer Science, 1993.

\bibitem{JonesSestoftSondergaard@LSC-89}
N.~D. Jones, P.~Ses{\-}toft, and H.~S{\o}n{\-}der{\-}gaard.
\newblock {MIX: a Self-applicable Partial Evaluator for experiments in Compiler
  Generation}.
\newblock {\em LISP and Symbolic Computation}, 2(1):9--50, 1989.

\bibitem{JorgensenLeuschel@PE-96}
J.~J{\o}rgensen and M.~Leuschel.
\newblock {Efficiently Generating Efficient Generating Extensions in Prolog}.
\newblock In O.~Danvy, R.~Gl\"{u}ck, and P.~Thiemann, editors, {\em Proceedings
  of the 1996 Dagstuhl Seminar on Partial Evaluation}, number 1110 in LNCS,
  pages 238--262, Schlo\ss\ Dagstuhl, February 1996. Springer-Verlag.

\bibitem{Komorowski@POPL-82}
J.~Komorowski.
\newblock {Partial Evaluation as a means for inferencing data structures in an
  Applicative Language: A Theory and Implementation in the case of Prolog}.
\newblock In {\em Proceedings of the ACM Conference on Principles of
  Programming Languages}, pages 255--267, Albuquerque, New Mexico, January
  1982. ACM.

\bibitem{Leuschel96:ecce-dppd}
M.~Leuschel.
\newblock The {\sc ecce} partial deduction system and the {\sc dppd} library of
  benchmarks.
\newblock Obtainable via {\tt http://www.cs.kuleuven.ac.be/\~{}lpai}, 1996.

\bibitem{LeuschelMartensDeSchreye@TOPLAS-98}
M.~Leuschel, B.~Martens, and D.~De~Schreye.
\newblock {Controlling Generalisation and Polyvariance in Partial Deduction of
  Normal Logic Programs}.
\newblock {\em ACM Trans. Prog. Lang. Syst.}, 20, 1998.
\newblock To Appear.

\bibitem{TabledUnfold@LOPSTR-97}
M.~Leuschel, B.~Martens, and K.~Sagonas.
\newblock {Preserving Termination of Tabled Logic Programs While Unfolding}.
\newblock In N.~Fuchs, editor, {\em Proceedings of LOPSTR'97: Logic Program
  Synthesis and Transformation}, LNCS, Leuven, Belgium, July 1997.

\bibitem{LindenstraussSagiv@ICLP-97}
N.~Lindenstrauss and Y.~Sagiv.
\newblock {Automatic Termination Analysis of Logic Programs}.
\newblock In L.~Naish, editor, {\em Proceedings of the Fourteenth International
  Conference on Logic Programming}, pages 63--77, Leuven, Belgium, July 1997.
  The MIT Press.

\bibitem{MarchioriTeusink@ILPS-95}
E.~Marchiori and F.~Teusink.
\newblock {Proving Termination of Logic Programs with Delay Declarations}.
\newblock In J.~W. Lloyd, editor, {\em Proceedings of the 1995 International
  Logic Programming Symposium}, pages 447--461, Portland, Oregon, December
  1995.

\bibitem{MarriottSondergaard@LOPLAS-93}
K.~Marriott and H.~S{\o}ndergaard.
\newblock {Precise and Efficient Groundness Analysis for Logic Programs}.
\newblock {\em ACM Letters on Progr. Lang. and Syst.}, 2(1--4):181--196,
  1993.

\bibitem{MartensDeSchreye@JLP-96}
B.~Martens and D.~De~Schreye.
\newblock {Automatic Finite Unfolding Using Well-Founded Measures}.
\newblock {\em The Journal of Logic Programming}, 28(2):89--146, August 1996.

\bibitem{Martin}
J.~Martin.
\newblock {Sonic Partial Deduction}.
\newblock Technical Report, 
 Dept. Elec. and Comp. Sc.,
  University of Southampton, January 1998.

\bibitem{MogensenBondorf:LOPSTR92}
T.~Mogensen and A.~Bondorf.
\newblock Logimix: A self-applicable partial evaluator for {P}rolog.
\newblock In K.-K. Lau and T.~Clement, editors, {\em {\em Logic Program
  Synthesis and Transformation.} Proceedings of LOPSTR'92}, pages 214--227.
  Springer-Verlag, 1992.

\bibitem{Naish92}
L.~Naish.
\newblock Coroutining and the Construction of Terminating Logic Programs.
\newblock Technical Report TR 92/5, Dept. Computer Science, University of
  Melbourne, 1992.

\bibitem{XSB@SIGMOD-94}
K.~Sagonas, T.~Swift, and D.~S. Warren.
\newblock {XSB as an Efficient Deductive Database Engine}.
\newblock In {\em Proceedings of the ACM SIGMOD International Conference on the
  Management of Data}, pages 442--453, Minneapolis, Minnesota, May 1994. ACM.

\bibitem{Sahlin@NGC-93}
D.~Sahlin.
\newblock {Mixtus: An Automatic Partial Evaluator for Full Prolog}.
\newblock {\em New Generation Computing}, 12(1):7--51, 1993.

\bibitem{SorensenGlueck@ILPS-95}
M.~H. S{\o}rensen and R.~Gl\"{u}ck.
\newblock {An Algorithm of Generalization in Positive Supercompilation}.
\newblock In J.~W. Lloyd, editor, {\em Proceedings of the 1995 International
  Logic Programming Symposium}, pages 465--479, Portland, Oregon, December
  1995.

\bibitem{Turchin@TOPLAS-86}
V.~F. Turchin.
\newblock {The Concept of a Supercompiler}.
\newblock {\em ACM Trans. Prog. Lang. Syst.}, 8(3):292--325, July 1986.

\bibitem{PropGAIA@JLP-95}
P.~{Van Hentenryck}, A.~Cortesi, and B.~{Le Charlier}.
\newblock {Evaluation of the Domain Prop}.
\newblock {\em Journal of Logic Programming}, 23(3):237--278, June 1995.

\end{thebibliography}

\end{small}


\appendix

\section{Specialised Programs generated by {\sc bta + logen}}

\subsection{Parser}

Original program:
\begin{footnotesize}
\begin{verbatim}
nont(X,T,R) :- t(a,T,V),nont(X,V,R).
nont(X,T,R) :- t(X,T,R).
t(X,[X|Es],Es).
\end{verbatim}
\end{footnotesize}

\noindent
Partial deduction query:
\begin{footnotesize}
\begin{verbatim}
nont(c,X,Y).
\end{verbatim}
\end{footnotesize}

\noindent
Specialised program (where {\tt\small nont(c,X,Y)}
 has been renamed to {\tt\small nont\_\_0(X,Y)}):
\begin{footnotesize}
\begin{verbatim}
nont__0([a|B],C) :- nont__0(B,C).
nont__0([c|D],D).
\end{verbatim}
\end{footnotesize}

\subsection{Liftsolve.app}

Original program:
\begin{footnotesize}
\begin{verbatim}
solve(GrRules,[]).
solve(GrRules,[NgH|NgT]) :-
        non_ground_member(term(clause,[NgH|NgBody]),GrRules),
        solve(GrRules,NgBody),
        solve(GrRules,NgT).
 
non_ground_member(NgX,[GrH|_GrT]) :-
        make_non_ground(GrH,NgX).
non_ground_member(NgX,[_GrH|GrT]) :-
        non_ground_member(NgX,GrT).
 
make_non_ground(G,NG) :- mng(G,NG,[],Sub).
 
mng(var(N),X,[],[sub(N,X)]).
mng(var(N),X,[sub(N,X)|T],[sub(N,X)|T]).
mng(var(N),X,[sub(M,Y)|T],[sub(M,Y)|T1]) :-
        N \== M, mng(var(N),X,T,T1).
mng(term(F,Args),term(F,IArgs),InSub,OutSub) :-
        l_mng(Args,IArgs,InSub,OutSub).
 
l_mng([],[],Sub,Sub).
l_mng([H|T],[IH|IT],InSub,OutSub) :-
        mng(H,IH,InSub,IntSub),
        l_mng(T,IT,IntSub,OutSub).
\end{verbatim}
\end{footnotesize}

\noindent
Partial deduction query:
\begin{footnotesize}
\begin{verbatim}
solve([term(clause,[term(app,[term(null,[]),var(l),var(l)])]),
       term(clause,[term(app,[term(cons,[var(h),var(x)]),var(y),
          term(cons,[var(h),var(z)])]),term(app,[var(x),var(y),var(z)])])],
        [term(app,[X1,X2,X3])]).
\end{verbatim}
\end{footnotesize}

\noindent
Specialised program:
\begin{footnotesize}
\begin{verbatim}
solve__0([]).
solve__0([term(app,[term(null,[]),B,B])|C]) :- 
  solve__0([]), solve__0(C).
solve__0([term(app,[term(cons,[D,E]),F,term(cons,[D,G])])|H]) :- 
  solve__0([term(app,[E,F,G])]), solve__0(H).
\end{verbatim}
\end{footnotesize}

\subsection{Liftsolve.app4}

Same original program as {\tt liftsolve.app}.

\noindent
Partial deduction query:
\begin{footnotesize}
\begin{verbatim}
solve([term(clause,[term(app,[term(null,[]),var(l),var(l)])]),
       term(clause,[term(app,[term(cons,[var(h),var(x)]),var(y),
                              term(cons,[var(h),var(z)])]),
                    term(app2,[var(x),var(y),var(z)])]),
       term(clause,[term(app2,[term(null,[]),var(l),var(l)])]),
       term(clause,[term(app2,[term(cons,[var(h),var(x)]),var(y),
                               term(cons,[var(h),var(z)])]),
                    term(app3,[var(x),var(y),var(z)])]),
       term(clause,[term(app3,[term(null,[]),var(l),var(l)])]),
       term(clause,[term(app3,[term(cons,[var(h),var(x)]),var(y),
                               term(cons,[var(h),var(z)])]),
                    term(app4,[var(x),var(y),var(z)])]),
       term(clause,[term(app4,[term(null,[]),var(l),var(l)])]),
       term(clause,[term(app4,[term(cons,[var(h),var(x)]),var(y),
                               term(cons,[var(h),var(z)])]),
                    term(app,[var(x),var(y),var(z)])])],
       [term(app,[_X,_Y,_Z])])
\end{verbatim}
\end{footnotesize}

\noindent
Specialised program:
\begin{footnotesize}
\begin{verbatim}
solve__0([]).
solve__0([term(app,[term(null,[]),B,B])|C]) :- 
  solve__0([]),solve__0(C).
solve__0([term(app,[term(cons,[D,E]),F,term(cons,[D,G])])|H]) :- 
  solve__0([term(app2,[E,F,G])]),solve__0(H).
solve__0([term(app2,[term(null,[]),I,I])|J]) :- 
  solve__0([]),solve__0(J).
solve__0([term(app2,[term(cons,[K,L]),M,term(cons,[K,N])])|O]) :- 
  solve__0([term(app3,[L,M,N])]),solve__0(O).
solve__0([term(app3,[term(null,[]),P,P])|Q]) :- 
  solve__0([]),solve__0(Q).
solve__0([term(app3,[term(cons,[R,S]),T,term(cons,[R,U])])|V]) :- 
  solve__0([term(app4,[S,T,U])]),solve__0(V).
solve__0([term(app4,[term(null,[]),W,W])|X]) :- 
  solve__0([]),solve__0(X).
solve__0([term(app4,[term(cons,[Y,Z]),A_1,term(cons,[Y,B_1])])|C_1]) :- 
  solve__0([term(app,[Z,A_1,B_1])]),solve__0(C_1).
\end{verbatim}
\end{footnotesize}

\subsection{Depth}

Original program:
\begin{footnotesize}
\begin{verbatim}
depth( true, 0 ).
depth( (_g1,_gs), _depth ) :-
    depth( _g1, _depth_g1 ),
    depth( _gs, _depth_gs ),
    max( _depth_g1, _depth_gs, _depth ).
depth( _goal, s(_depth) ) :-
    prog_clause( _goal, _body ),
    depth( _body, _depth ).
\end{verbatim}
\end{footnotesize}

\noindent
Partial deduction query:
\begin{footnotesize}
\begin{verbatim}
depth(member(X,[a,b,c,m,d,e,m,f,g,m,i,j]),D).
\end{verbatim}
\end{footnotesize}

\noindent
Specialised program:
\begin{footnotesize}
\begin{verbatim}
depth__0(true,0).
depth__0(member(B,C),s(D)) :- 
  depth__0(append(E,[B|F],C),D).
depth__0(append([],G,G),s(H)) :- 
  depth__0(true,H).
depth__0(append([I|J],K,[I|L]),s(M)) :- 
  depth__0(append(J,K,L),M).
\end{verbatim}
\end{footnotesize}

\subsection{Match.Kmp}

Original program:
\begin{footnotesize}
\begin{verbatim}
match(Pat,T) :- match1(Pat,T,Pat,T).
 
match1([],Ts,P,T).
match1([A|Ps],[B|Ts],P,[X|T]) :-
        A\==B,match1(P,T,P,T).
match1([A|Ps],[A|Ts],P,T) :-
        match1(Ps,Ts,P,T).
\end{verbatim}
\end{footnotesize}

\noindent
Partial deduction query:
\begin{footnotesize}
\begin{verbatim}
match([a,a,b],R).
\end{verbatim}
\end{footnotesize}

\noindent
Specialised program:
\begin{footnotesize}
\begin{verbatim}
match1__4(B,C).
match1__3([B|C],[D|E]) :- \==(b,B), match1__1(E,E).
match1__3([b|F],G) :- match1__4(F,G).
match1__2([B|C],[D|E]) :- \==(a,B), match1__1(E,E).
match1__2([a|F],G) :- match1__3(F,G).
match1__1([B|C],[D|E]) :- \==(a,B), match1__1(E,E).
match1__1([a|F],G) :- match1__2(F,G).
match__0(B) :- match1__1(B,B).
\end{verbatim}
\end{footnotesize}

\subsection{Regexp.r1}

Original program:
\begin{footnotesize}
\begin{verbatim}
generate(empty,T,T).
generate(char(X),[X|T],T).
generate(or(X,Y),H,T) :- generate(X,H,T).
generate(or(X,Y),H,T) :- generate(Y,H,T).
generate(cat(X,Y),H,T) :- generate(X,H,T1), generate(Y,T1,T).
generate(star(X),T,T).
generate(star(X),H,T) :- generate(X,H,T1), generate(star(X),T1,T).
\end{verbatim}
\end{footnotesize}

\noindent
Partial deduction query:
\begin{footnotesize}
\begin{verbatim}
generate(cat(star(or(char(a),char(b))),
         cat(char(a),cat(char(a),char(b)))),X1,[])
\end{verbatim}
\end{footnotesize}

\noindent
Specialised program:
\begin{footnotesize}
\begin{verbatim}
generate__3([a|B],B).
generate__4([b|B],B).
generate__2(B,C) :- generate__3(B,C).
generate__2(D,E) :- generate__4(D,E).
generate__1(B,B).
generate__1(C,D) :- generate__2(C,E), generate__1(E,D).
generate__6(B,C) :- generate__3(B,D), generate__4(D,C).
generate__5(B,C) :- generate__3(B,D), generate__6(D,C).
generate__0(B,C) :- generate__1(B,D), generate__5(D,C).
\end{verbatim}
\end{footnotesize}

\end{document}